\documentclass[12pt]{article}
\usepackage{graphicx}
\usepackage{amsopn}
\usepackage{amsfonts}
\usepackage{latexsym}
\usepackage{amsmath,bm,amssymb}
\usepackage{authblk}

\begin{document}

\title{Pair correlation function of vorticity in a coherent vortex}

\author{I.V.Kolokolov$^1 \, ^2$, V.V.Lebedev$^1\, ^2$, M.M.Tumakova$^1\, ^3$}

\affil{$^1$ Landau Institute for Theoretical Physics, RAS, \\
142432, Chernogolovka, Moscow region, Russia; \\
$2$ National Research  University Higher School of Economics, \\
101000, Myasnitskaya ul. 20, Moscow, Russia.\\
$^3$ National Research  University Higher School of Economics, 190008,
 St.Petersburg, Russia}

\maketitle

\begin{abstract}

We study the correlations of vorticity fluctuations inside a coherent vortex resulting from the inverse energy cascade in two-dimensional turbulence. The presence of a coherent flow, which is a differential rotation, suppresses small-scale fluctuations of the flow, which are created by an external force, and lead to the fact that these fluctuations can be considered as non-interacting and, therefore, examined in a linear approximation. We calculate the pair correlation function of vorticity and demonstrate that it has a power-law behavior both in space and in time. The obtained results allow us to start a systematic study of the effects associated with the nonlinear interaction of fluctuations, which play an essential role on the periphery of a coherent vortex. Our results are also applicable to the statistics of a passive scalar in a strong shear flow.

\end{abstract}

As shown in the works \cite{67Kra,68Lei,69Bat}, the main feature of a two-dimensional turbulent flow excited by an external force (pumping) is the presence of the inverse energy cascade, which is formed along with the direct enstrophy cascade. A description of the causes of the inverse cascade and of its properties is contained in Ref. \cite{KraM}. An overview of the current state of two-dimensional turbulence research can be found in Ref. \cite{boffetta2012two}).

After the pumping (an external force that excites turbulence) is turned on, due to the inverse cascade the spatial scale of velocity fluctuations increases as time goes until this growth is stopped either by the bottom friction or by the size of the box. In the latter case, the final stage of evolution may be a system of coherent vortices. Experimentally coherent vortices were observed in \cite{ShaF,FilLevchOrlov17}, in numerical modeling they were observed in \cite{Smith2, chertkov2007dynamics,laurie2014universal}. In the last work, a flat velocity profile of a coherent vortex is established and arguments explaining its occurrence are given. Further analytical studies of \cite{KL15,kolokolov2016structure,kolokolov2016velocity,frishman17,kolokolov2017static} confirmed the universality of the flat profile and linked its existence with the passive (quasi-linear) nature of fluctuations inside the coherent vortex. In Ref. \cite{KL20} it was shown that coherent vortices cannot be formed when the coefficient of friction on the bottom exceeds a certain threshold, in Ref. \cite{DFKL22} this prediction was verified by direct numerical modeling.

From the point of view of energy balance, the coherent large-scale vortices arising in two-dimensional turbulent systems are supported by small-scale fluctuations excited by an external force. These fluctuations, in turn, evolve in a large-scale velocity field. Their amplitudes are suppressed inside the coherent vortex and in the leading approximation, the nonlinear interaction of small-scale fluctuations can be neglected. The correlation functions of such fluctuations are strongly anisotropic. In this paper, we calculate the pair correlation function of the fluctuation component of the vorticity field inside a coherent vortex in the passive approximation. Such a function is directly measurable in laboratory and numerical experiments, and comparing it with theoretical prediction can serve as an indicator of the applicability of the entire approach. In addition, the analysis is the first necessary step in the study of the role of nonlinear fluctuation effects that determine the statistical properties of the flow outside the coherent vortex and its size.

In the reference frame attached to the center of the coherent vortex, the flow is isotropic on average, that is, the mean velocity of the vortex $\bm V$ has only the polar component $U(r)$, where $r$ is the distance to the observation point from the center of the vortex. The total velocity of the fluid flow inside the vortex can be represented as the sum of $\bm V+\bm v$, where $\bm v$ is the fluctuation component of the velocity. A similar decomposition, $\Omega+\varpi$, is true for vorticity, where $\Omega=\partial_r U+U/r$, $\varpi=\nabla\times\bm v$, . The fluctuation components $\bm v$, $\varpi$ depend on the time $t$, on the distance $r$ and on the polar angle $\varphi$.

We proceed from the two-dimensional Navier-Stokes equation, in which, along with the viscous term, we take into account the bottom friction. After decomposing the flow into a coherent and fluctuation components, we find from the Navier-Stokes equation the following equation for fluctuation vorticity $\varpi$
 \begin{equation}
 \partial_t\varpi+(U/r)\partial_\varphi\varpi+v_r\partial_r\Omega
 +\nabla\left(\bm v \varpi-\langle \bm v \varpi \rangle\right)
 =\phi-\alpha \varpi +\nu \nabla^2 \varpi.
 \label{vact}
 \end{equation}
Here $\phi=\nabla\times \bm f$ is curl of the external force $\bm f$ (per unit mass), $v_r$ is the radial component of the fluctuational velocity $\bm v$, $\alpha$ is the bottom friction coefficient, and $\nu$ is the kinematic viscosity coefficient.

We assume that the external force $\bm f$, exciting turbulence, is chaotically varying in time and having a finite correlation length in space. We use the model where the force $\bm f$ is short-correlated in time. In this case, its statistical properties are unambiguously characterized by the pair correlation function of $\bm f$. It is convenient for us to work in terms of the correlation function of the curl of the external force $\phi$
\begin{equation}
\langle\phi(t_1,\bm x)\phi(t_2,\bm y)\rangle
=-\epsilon\delta(t_1-t_2) \nabla^2 \Xi(\bm x-\bm y),
\label{basic2}
\end{equation}
where we assumed that the pumping statistical properties are homogeneous in time and space. Laplacian $\nabla^2$ in the right hand side of Eq. (\ref{basic2}) is related to the fact that $\phi$ is the curl of the external force, $\epsilon$ is the pumping power per unit mass, and $\Xi(\bm x)$ is a dimensionless function satisfying the condition $\Xi(0)=$1. We believe that $\Xi(\bm x)$ decreases rapidly as its argument grows on scales larger than some length $k_f^{-1}$, which is the correlation length of the pumping. We assume that the length of $k_f^{-1}$ is small compared to the distance $r$ from the observation point to the center of the vortex, $k_f r\gg 1$.

Outside the core of the coherent vortex (where viscosity plays the main role and solid-state rotation is realized), its average polar velocity $U$ has a flat profile \cite{laurie2014universal,kolokolov2017static}, which corresponds to differential rotation. Locally (on scales less than the distance to the center of the vortex) we are dealing with a shear flow. As shown in the works \cite{kolokolov2016structure,kolokolov2016velocity}, the shear rate of the mean velocity, $\Sigma=r\partial_r(U/r)$, is much larger than the Kolmogorov nonlinear frequency of interaction of fluctuations on the pumping length $(\epsilon k_f^2)^{2/3}$. Therefore, the nonlinear terms in Eq. (\ref{vact}) for the  fluctuation vorticity can be neglected in the leading approximation. Thus, the fluctuations on the background of a coherent vortex can be analyzed in a passive (quasi-linear) approximation.

Note that for the formation of a two-dimensional turbulent state and of a coherent vortex, the dissipative parameters must be sufficiently small. In particular, if a coherent vortex has formed, then the inequality holds inside it
\begin{equation}
\Sigma\gg \nu k_f^2.
\label{sigmabig}
\end{equation}
The inequality (\ref{sigmabig}) holds true everywhere inside the vortex except for a narrow area near the center of the vortex, where $\Sigma\to0$. Despite the relative weakness of the dissipative terms, it is impossible to neglect them when analyzing a coherent vortex, since it is the viscosity and the bottom friction that determine the amplitudes of the mean (coherent) velocity of the vortex and the fluctuations on its background.

Let us consider a small neighborhood of a circle of radius $R$ centered in the center of the vortex. The isotropy of its mean flow makes it possible to exclude homogeneous rotation with an angular frequency of $U(R)/R$ by switching to a frame of reference rotating with this angular velocity. When solving the equation (\ref{vact}), we limit ourselves to the values of $r$ close to $R$. Then the variables $x_1=R\varphi$ and $x_2=r-R$ can be treated local Cartesian coordinates. The inequality $k_f R\gg 1$ implies, that we can neglect the term $v_r\partial_r \Omega$ in the equation (\ref{vact}) and, consequently, to use the approximation of the shear flow. The equation of evolution of fluctuation vorticity thus takes the form
\begin{equation}
\hat{\cal L}\varpi=\phi, \quad
\hat{\cal L}=\partial_t
+\Sigma x_2 \frac{\partial}{\partial x_1}
+\alpha -\nu \nabla^2.
\label{basic1}
\end{equation}
Here $\Sigma$ is the shear rate, $\Sigma=r \partial_r(U/r)$, taken at $r=R$.

The solution of the equation (\ref{basic1}) is written as
\begin{equation}
\varpi(t,\bm x)=
\int dt_1 d^2 y\, G(t-t_1,\bm x, \bm y)\phi(t_1,\bm y).
\label{greeni}
\end{equation}
Here $G(t,\bm x, \bm z)$ is the Green function of the evolutionary operator $\hat{\cal L}$ (\ref{basic1}), satisfying the equation
\begin{equation}
\hat{\cal L}G(t,\bm x, \bm z) =\delta(t)\delta(\bm x-\bm z).
\label{greenf}
\end{equation}
Due to causality, the Green function $G(t)$ is equal to zero if $t<0$.

For the spatial Fourier component of the Green function, we obtain from (\ref{greenf}) a first-order differential equation, which is easily solved by the method of characteristics. After that, the inverse Fourier transform gives
\begin{eqnarray}
G(t,\bm x,\bm y)=\theta(t)
\frac{\sqrt 3\,\exp(-\alpha t)}{2\pi \nu \Sigma t^2}
\exp\left\{-\frac{3\Delta(t,\bm x,\bm y)}{\nu \Sigma^2 t^3}\right\},
\label{gref4} \\
\Delta(t,\bm x,\bm y)=(x_1-y_1-\Sigma t y_2)^2
\nonumber \\
-\Sigma t (x_1-y_1-\Sigma ty_2)(x_2-y_2)
+\frac{1}{3}\Sigma^2 t^2 (x_2-y_2)^2,
\label{gref5}
\end{eqnarray}
where $\theta(t)$ is Heaviside step function. Note that the bottom friction enters the expression (\ref{gref4}) solely through the factor $\exp(-\alpha t)$.

The expression (\ref{gref4}) demonstrates, that the evolution of the vorticity blob of size $k_f^{-1}$ produced by pumping $\phi$ is characterized by different times for different directions. Across the shear flow, the characteristic evolution time is determined by the viscosity, it is equal to $\tau_\nu=(\nu k_f^2)^{-1}$. Along the shear flow, the characteristic time of evolution is equal to
\begin{equation}
\tau_\star =\left(\Sigma^2 \nu k_f^2\right)^{-1/3}.
\label{taustar}
\end{equation}
By virtue of the condition (\ref{sigmabig}), the inequalities are valid $\Sigma \tau_\star \gg 1$ and $\tau_\nu\gg \tau_\star$. As it demonstrated in the works \cite{KL20,DFKL22}, a condition for the emergence of a coherent vortex is $\alpha \lesssim \nu k_f^2$. Therefore one finds $\alpha \tau_\star \ll 1$, as a consequence of the same condition (\ref{sigmabig}).

The time $\tau_\star$ has the following origin: Langevin forces providing viscous diffusion lead to the fact that Lagrangian trajectories have an angular spread $\theta_\star \sim \left(\nu k_f^2/\Sigma\right)^{1/3}$. Such angle spread appears, for example, in the problem of polymer stretching in a shear flow with a small random component \cite{poly05}. As a result, the shear flow generates a velocity spread on the pumping scale $\Delta v \sim \theta_\star  k_f^{-1}\Sigma$, which gives the effective time of deformation of the blob (\ref{taustar}). This interpretation is consistent with the results of the study of the decay of the initial blob of the passive scalar, given in Ref. \cite{souzy}, where it was shown that the characteristic decay time is estimated as $\left(D \Sigma^2\right)^{-1/3}l^{2/3}$, where $l$ is the size of the blob and $D$ is the diffusion coefficient of the scalar.

Let's study the pair correlation function $F$ of vorticity $\varpi$ in conditions of statistical equilibrium. It depends only on the time difference (due to the assumed uniformity in time), but at a non-zero time difference, it depends not only on the coordinate difference, since the shear flow violates the spatial uniformity of the system. Therefore, we are dealing with a function
\begin{equation}
F(t,\bm x,\bm y)
= \langle \varpi(t,\bm x) \varpi(0,\bm y)\rangle,
\label{paircr}
\end{equation}
where angle brackets mean averaging over the pumping statistics. The pumping has the correlation length $k_f^{-1}$, while the correlation function (\ref{paircr}) has a complex spatiotemporal behavior.

Utilizing the expression (\ref{greeni}) and averaging the product $\varpi(t,\bm x) \varpi(0,\bm y)$ using the relation (\ref{basic2}), we find the following expression
\begin{eqnarray}
F(t,\bm x,\bm y)
=-\epsilon\int_0^\infty dt_2 \int d^2s\, d^2z\,
G(t+t_2,\bm x,\bm s) G(t_2,\bm y, \bm z)\nabla^2 \Xi(\bm s-\bm z).
\label{gref78}
\end{eqnarray}
The expression (\ref{gref78}) allows a detailed analysis of the pair correlation function of vorticity. We will be interested in the power asymptotics of the vorticity correlation function (\ref{paircr}). These asymptotics are determined by the evolution operator (\ref{basic1}) and are therefore universal, that is, they do not depend on the pumping details.

We start by analyzing the second moment $\langle \varpi^2\rangle=F(0,\bm 0,\bm 0)$. Then the characteristic value $s_1-z_1\sim k_f^{-1}$ in the integral  (\ref{gref78}) is determined by the puping. The characteristic values of the remaining quantities in the integral (\ref{gref78}) are determined by the structure of the Green function (\ref{gref4}). One can demonstrate that
\begin{equation}
t_2\sim \tau_\star, \
s_1, z_1\sim k_f^{-1}, \
s_2,z_2\sim \sqrt{\nu \tau_\star}\ll k_f^{-1}.
\nonumber
\end{equation}
Using these characteristic values, we find from the integral (\ref{gref78})
\begin{equation}
\langle \varpi^2\rangle
=F(0,\bm 0,\bm 0)
\sim \epsilon k_f^2\tau_\star.
\label{asym1}
\end{equation}
One can say that $\tau_\star$ makes sense of the effective pumping time. Note that when obtaining (\ref{asym1}), we neglected the factor $\exp(\alpha t)$ in Eq. (\ref{gref4}), which is possible due to $\alpha \tau_\star\ll1$.

Next, we consider the single-point correlation function $F(t, \bm 0,\bm 0)$ assuming $t/\tau_\star \gg 1$. Then the characteristic time $t_2$ in the integral  (\ref{gref78}) is determined by the estimate $t_2\sim t$. The quantities $\bm s,\bm z$ are determined by the estimates
\begin{equation}
s_1- z_1\sim k_f^{-1}, \
s_1+z_1 \sim \sqrt{\nu \Sigma^2 t^3}, \
s_2\sim z_2 \sim \sqrt{\nu t}.
\nonumber
\end{equation}
Using the characteristic values, we find from Eq. (\ref{gref78})
\begin{equation}
F(t, \bm 0,\bm 0) \sim
\epsilon  k_f^2 \tau_\star \sqrt{\tau_\star /t}.
\label{asym2}
\end{equation}
The estimate (\ref{asym2}) implies the inequality $k_f s_2 \ll1$, that leads to $t\ll \tau_\nu=(\nu k_f^2)^{-1}$. Otherwise, at $t\gg \tau_\nu$, limitations in the integration over $s_2,z_2$ related to pumping become significant, and the value of $F(t, \bm 0,\bm 0)$ diminishes fast as $t$ grows. When calculating (\ref{asym2}), we neglected the factor $\exp(\alpha t)$ in Eq. (\ref{gref4}), that is correct at $\alpha t\ll1$. Note that due to the relation $\alpha \lesssim \nu k_f^2$ the restriction $t\gg \tau_\nu$ is stronger than one related to $\alpha$.

Spatial correlations of the field $\varpi$ turn out to be substantially anisotropic, which is a consequence of the presence of a strong shear flow. Therefore, we separately investigate the behavior of the correlation function $F$ for coordinates measured along and across the shear velocity. To establish the nature of spatial correlations, we investigate the expression for the simultaneous correlation function $F(0,\bm x,\bm y)$, which depends solely on the difference $\bm x-\bm y$.

Let us analyse the behaviour of  $F$ when the obser-
vation points are distant along the direction of the
shear flow, $F(0, x_1,0,y_1,0)$. Its value differs significantly on the second moment if $k_f|x_1-y_1|\gg 1$. For this case the following estimations hold:
\begin{eqnarray}
s_1-z_1 \sim k_f^{-1}, \quad
s_1+z_1\sim |x_1-y_1|,
\nonumber \\
t_2\sim \tau_\star  (k_f |x_1-y_1|)^{2/3}, \quad
s_2\sim z_2 \sim \sqrt{\nu t_2}.
\nonumber
\end{eqnarray}
Using the given characteristic values, we find from the representation (\ref{gref78})
\begin{equation}
F(0, x_1,0,y_1,0)\sim
\frac{\epsilon k_f^2\tau_\star}{(k_f|x_1-y_1|)^{1/3}}.
\label{asym3}
\end{equation}
The estimate (\ref{asym3}) implies the inequality $k_f s_2 \ll1$, that leads to the resriction $t_2\ll \tau_\nu$. It is rewritten as the inequality $k_f |x_1-y_1|\ll (\tau_\nu/ \tau_\star)^{3/2}$, which determines the scope of applicability of the asymptotic (\ref{asym3}). Under the opposite condition, the correlation function decreases rapidly as $|x_1-y_1|$ grows. At establishing (\ref{asym3}) we neglected the factor $\exp(\alpha t)$ in Eq. (\ref{gref4}), that leads to the restriction $\alpha t_2\ll1$. Note that due to the relation $\alpha \lesssim \nu k_f^2$ the restriction $t_2\ll \tau_\nu$ is stronger than one related to $\alpha$.

Let's analyze the expression for the simultaneous correlation function $F$ for the configuration when the observation points are spaced across the direction of the shear flow, that is $F(0, 0,x_2,0,y_2)$. Its value differs essentially from the second moment $\langle \varpi^2\rangle$, if the points are spaced by a distance that satisfies $k_f|x_2-y_2|\gg (\nu k_f^2/\Sigma)^{1/3}$. Then the following estimates are valid
\begin{eqnarray}
s_2\sim z_2 \sim |x_2-y_2| \sim \sqrt{\nu t_2},
\nonumber \\
s_1-z_1 \sim k_f^{-1}, \quad
s_1+z_1 \sim (\nu \Sigma^2 t_2^3)^{1/2}
\sim (\Sigma/\nu)   |x_2-y_2|^3.
\nonumber
\end{eqnarray}
The function $F$ passes to the asymptotic regime where $s_1+z_1$ becomes larger than $k_f^{-1}$, the condition leads to the above criterion $k_f|x_2-y_2|\gg (\nu k_f^2/\Sigma)^{1/3}$. Using the given characteristic values, we find from the representation (\ref{gref78})
\begin{equation}
F(0,0, x_2, 0,y_2)\sim
\frac{\epsilon k_f}{\Sigma |x_2-y_2|}.
\label{omat}
\end{equation}
The asymptotic behavior (\ref{omat}) is realized at $k_f|x_2-y_2|\ll1$. Otherwise, the integration over $s_2-z_2$ is limited by the pump correlation function, and $F$ decreases rapidly as $|x_2-y_2|$ grows. Note that the largest value of $t_2$, realized where $k_f|x_2-y_2|\sim 1$, is $\tau_\star$.Therefore, when calculating (\ref{omat}), we neglected the factor $\exp(\alpha t)$ in Eq. (\ref{gref4}) due to $\alpha t_\star\ll1$.

We have established that the pair correlation function of vorticity inside a coherent vortex, which arises as a result of an inverse cascade in two-dimensional turbulence, has a power-law behavior in a certain regions of spatial lengths and times. This power-law behavior means a strong correlation of the vorticity fluctuations in the domain of power-law behavior, which can significantly affect nonlinear effects. Our approach has a broader scope of applicability, since it determines the correlations of a passive scalar in a strong shear flow. Our results can be directly compared with the results of laboratory experiments and numerical modeling.

We are grateful to S.S.Vergeles for numerous discussions.
I.V.K. and V.V.L.'s work was supported within the framework of the scientific program of the National Center for Physics and Mathematics, M.M.T.'s work was supported by the RFBR grant 19-32-60065 and "Basis" foundation, V.V.L. thanks also for the support of the grant of the Ministry of Education and Science of the Russian Federation, project No. 075-15-2022-1099.

\end{document}